\documentclass{styles/svproc}
\usepackage{url}

\usepackage{multirow}
\usepackage{subcaption}
\usepackage{amsmath}
\usepackage{amsfonts}
\usepackage{amssymb}
\usepackage[ruled,vlined]{algorithm2e}
\usepackage{tcolorbox}
\usepackage{makecell}
\usepackage{ctable}

\usepackage{diagbox}

\newcommand{\draft}[1]{}

\newcommand{\todo}[2]{}
\newcommand{\largehline}{\specialrule{.16em}{.3em}{.3em}}
\newcommand{\mediumhline}{\specialrule{.10em}{.2em}{.2em}}

\DeclareMathOperator{\task}{\mathcal{T}}

\begin{document}\begin{sloppypar}
\title{Continual Learning of Long Topic Sequences in Neural Information Retrieval}

\author{Thomas Gerald \and Laure Soulier}

\institute{CNRS-ISIR, Sorbonne University, Paris, France \\ \{gerald, soulier\}@isir.upmc.fr}

\maketitle

\begin{abstract}

In information retrieval (\textit{IR}) systems, trends and users' interests may change over time, altering either the distribution of requests or contents to be recommended. Since neural ranking approaches heavily depend on the training data, it is crucial to understand the transfer capacity of recent \textit{IR} approaches to address new domains in the long term.  In this paper, we first propose a dataset based upon the MSMarco corpus aiming at modeling a long stream of topics as well as \textit{IR} property-driven controlled settings. We then in-depth analyze the ability of recent neural \textit{IR} models while continually learning those streams.
Our empirical study highlights in which particular cases catastrophic forgetting occurs (e.g., level of similarity between tasks, peculiarities on text length, and ways of learning models) to provide future directions in terms of model design.  
\vspace{-0.4cm}

\end{abstract}

\keywords{Continual learning, Information retrieval, Neural ranking models}

\section{Introduction}
\vspace{-0.4cm}
% Proposition Laure
The information Retrieval (IR) field has seen a keen interest in neural approaches   these last years \cite{drmn2018mcdonald,dense2020Karpukhin,cedr2019macaveney,pacrr2017hui} thanks to recent advances in semantic and language understanding.
However, these approaches are heavily data dependent, often leading to specialization for a certain type of corpus \cite{MitraC18,OnalZARKBDCKMAB18}.
If document retrieval remains a core task, many challenges revolve around, such as news detection \cite{newsDetection}, question answering \cite{Yang-finetuning} or conversational search \cite{gao-conversational}. In all these tasks, users' needs or document content might evolve through time; leading to evolving queries and/or documents and shifting the topic distribution at the inference step \cite{Fei2014,McCreadieDAMLMO14,newsDetection}. It is, therefore, crucial to understand whether \textit{IR} models are able to change their ranking abilities to new topics/trends, but also to be still able to perform on previous topics/trends if these ones remain up to date.
Accumulating and preserving knowledge is thus an important feature in \textit{IR}, allowing to continuously adapt to new domains or corpora while still being effective on the old ones. This requirement refers to an emerging research field called \textit{Continual learning} \cite{overcoming2017kirpatrick,icarl2017rebuffi,efficient2020veniat}. In practice, continual learning proposes to learn all tasks sequentially by guaranteeing that previous knowledge does not deteriorate through the learning process; this phenomenon is called \textit{catastrophic forgetting}. To solve this issue, one might consider multi-task learning \cite{multitask} in which models learn together all the sets of tasks. Another approach would consists in learning a model for each task, but, in this case, the knowledge is not transferred between previous and current tasks.
These two last settings are not always realistic in \textit{IR}, since they consider that all tasks are available at the training step. In practice,  content and users' needs may evolve throughout the time \cite{McCreadieDAMLMO14,Fei2014}.

To the best of our knowledge, only one previous work has addressed the continual learning setting in \textit{IR} \cite{studying2021lovon}, highlighting the small weakness of the studied neural models to slightly forget knowledge over time. However this work has two limitations: 1) it only considers few tasks in the stream (2 or 3 successive datasets) and does not allow to exhibit neural model abilities in the more realistic scenario of long-term topic sequences (i.e., a larger number of users and topics implying evolving information needs/trends).  2) Although authors in \cite{studying2021lovon} use datasets of different domains, there is no control of stream properties (e.g., language shift \cite{languagedrift,Fei2014}, information update \cite{McCreadieDAMLMO14}) allowing to correlate the observed results with \textit{IR} realistic settings, as done in \cite{efficient2020veniat} for classification tasks.
 
The objective of this paper is thus to provide a low-level analysis of the learning behavior of neural ranking models through a continual setting considering long sequences and \textit{IR}-driven controlled topic sequences. 
In this aim, we propose to study different neural ranking models and to evaluate their abilities to preserve knowledge. To this end, we consider neural rankers successively fined tuned on each task of the sequence.
More particularly, our contribution is threefold:
\vspace{-0.2cm}
\begin{itemize}
    \item[$\bullet$] We design a corpus derived from the  MSMarco Passage ranking dataset \cite{NguyenRSGTMD16} to address long sequences of topics for continual learning and \textit{IR}-driven controlled topic sequences (Section 4).
    \item[$\bullet$] We compare the different neural ranking models in a long-term continual \textit{IR} setting (Section 5.1) and the controlled settings (Section 5.3).
    \item[$\bullet$] We in-depth investigate the impact of task similarity level in the continual setting on the learning behavior of neural ranking models (Section 5.2).

\end{itemize}

\vspace{-0.4cm}
\section{Related Works}
\vspace{-0.2cm}

\paragraph{Neural Information Retrieval.}

Deep learning algorithms have been introduced in \textit{IR} to learn representations of tokens/words/texts as vectors and compare query and document representations   \cite{GuoFAC16,drmn2018mcdonald,knrm2017xiong,pacrr2017hui,copacrr2018hui,convknrm2018dai}. 
With the advance of sequence-to-sequence models, semantic matching models have grown in popularity, particularly due to the design of new mechanisms such the well-known self-attention in transformer networks \cite{attention2017vaswani} or language models such as Bert \cite{bert2019devlin}.

Many \textit{IR} approaches benefit from those advances as \textit{CEDR} \cite{cedr2019macaveney} that combines a Bert language model with relevance matching approaches including KNRM \cite{knrm2017xiong} and PACRR \cite{pacrr2017hui}. Moreover, recent works addressed ranking with sequence-to-sequence transformers based approach as the \textit{Mono-T5} model \cite{document2020nogueira} for re-ranking documents returned by a BM25 ranker. 
%One downside of the latter architecture is the processing time: transformer self-attention processing is quadratic regarding the size of inputs. To benefit from the performance of those architectures performances while ensuring fast ranking, many approaches rely on the re-ranking principle. 
Using a weak initial ranker such as BM25 may be the bottleneck of reaching higher performances, some approaches  are thus reconsidering dense retrieval \cite{dense2020Karpukhin,colbert2020khattab,FormalPC21,ZhaoLL21}. All these models are data-dependent, relying on word/topic/query distribution in the training dataset and their application to new domains is not always straightforward \cite{MitraC18,OnalZARKBDCKMAB18}. While previous works addressed this issue by leveraging for instance fine-tuning techniques \cite{MaSA21,Yang-finetuning}, one can wonder whether these models are still effective on the word/topic/query distribution of the training dataset. This condition is particularly crucial for open-domain \textit{IR} systems (e.g., public search engines or future conversational search systems) since they should be able to face multiple users and solve both persistent information needs and event-related ones. 

\vspace{-0.2cm}
\paragraph{Continual Learning.}

Continual learning generally defines the setting in which a model is trained consecutively on a sequence of tasks and need to adapt itself to new encountered tasks. One main issue of continual learning is that models need to acquire knowledge throughout the sequence without forgetting the knowledge learnt on previous tasks (\textit{catastrophic forgetting}). 
To solve the catastrophic forgetting issue, three main categories can be outlined \cite{continual2019delange}. 
First, regularisation approaches continually learn to address new tasks using soft or hard preservation of weights \cite{overcoming2017kirpatrick,8107520,wiese-etal-2017-neural-domain}. For instance, the \textit{Elastic Weight Consolidation} model \cite{overcoming2017kirpatrick} softly updates weights for a new task according to their importance in the previous one.
Second, replay approaches \cite{icarl2017rebuffi,Ashar20,journals/corr/abs-1906-01076} (or \textit{rehearsal approaches}), replay examples of previous tasks while training the model on a new one. 
Third, architecture-based approaches  \cite{cai-etal-2019-adaptive,learn2019li,efficient2020veniat} rely on the decomposition of the inference function.
%, such that some parts are shared or not during inference. 
For instance, new approaches leveraging techniques of neural architecture search \cite{learn2019li,efficient2020veniat} have been proposed. 

 Recently some works have addressed the continual learning setting for \textit{NLP}  tasks.
 %For instance, the 
 %LAMOL model  \cite{lamol2020sun}  a continual language model based on the rehearsal approach. Other applications can also be found for 
 %such conversational systems \cite{Lee2017} or translation tasks \cite{GarciaCPF21}. 
 LAMOL \cite{lamol2020sun} for continual language modelling, \cite{Lee2017}  for conversational systems or \cite{GarciaCPF21} for translations tasks.
 While it exists \textit{IR} approaches to perform on different domains such as using batch balanced topics \cite{DBLP:conf/sigir/HofstatterLYLH21}, at the best of our knowledge, only one study addresses \textit{IR} in the continual setting \cite{studying2021lovon}, comparing neural ranking models on three successive tasks (MSMarco, TREC-Microblog, and TREC CORD19). Our work follows this line by providing an analysis of the behavior of neural ranking models on longer sequences of topics. We also design \textit{IR}-driven controlled sequences to highlight to what extent neural models face \textit{IR}-specific divergences, such as language drift or documents collection update.

\vspace{-0.2cm}
\section{Research design for continual learning in \textit{IR}}
\vspace{-0.4cm}
\label{section:continual-ir}
We address in this paper the following research questions aiming at analyzing the resilience of \textit{IR} models to catastrophic forgetting:\\
 $\bullet$ \textbf{RQ1:} How to design a sequence of tasks for continual learning in \textit{IR}? \\
$\bullet$ \textbf{RQ2:} What are the performance of neural ranking models while learning long sequences of topics? Can we perceive signals of catastrophic forgetting?\\
 $\bullet$ \textbf{RQ3:}  Does the similarity level of tasks in the sequence impact the model effectiveness and their robustness to catastrophic forgetting? \\
 $\bullet$ \textbf{RQ4:} How do neural ranking models adapt themselves to queries or documents distribution shifts?
\vspace{-0.2cm}
\subsection{Continual learning setting and metrics}
\vspace{-0.2cm}
We propose a continual learning setting based on long sequences. The latter consists in fine-tuning a model on different tasks successively. Following \cite{studying2021lovon}, we instantiate tasks by topics/domains, but we rather focus on long sequences of tasks with the perspective that
%, for instance, 
such setting can be connected with long-term trends/changes of user interests. 
In practice, we consider a sequence 
%$S$
of $n$ tasks $S =\{\task_1, \ldots, \task_i, \ldots, \task_n\}$, each task $\task_i$ corresponds to a set of queries and their associated relevant documents. We suppose that each task relies on different properties or distributions as in \cite{Pan2010}.
Neural ranking models are successively fine-tuned over the long sequence $S$ of topics. The objective is to track each task and evaluate each of them at different timestep of the sequence (i.e., after the successive fine-tuning) to measure the model's abilities to adapt to new tasks and their resilience to catastrophic forgetting.

 In practice, we propose to track in each sequence a subset of 5 randomly selected tasks (tracking whole tasks throughout the whole sequence is too computationally expensive). For each of these tasks, we will measure at each step of the topic sequence
 %different ranking metrics: the MAP@K and
 the MRR@K. 
To measure the catastrophic forgetting $mf$ for a given task $\task_i$ at a training step $\theta_j$ (associated to task $\task_j$), we identify the maximum value obtained by the model along the sequence $S$ and compare its performances at each training step $\theta_j$ with the maximum value:  
%Formally, let be $score(i, \theta_j)$ referring to a ranking metric for the task $\task_i$ using the model obtained training the $j^{th}$ task $\task_j$ in the sequence. 
%We define a forgetting score $mf$ for measuring forgetting based on the maximum reference and compare the maximum score to the score of $\task_i$ with the weights obtained after training $\task_j$:
\vspace{-0.2cm}
\begin{equation}
    mf(i, \theta_j) =  \left( \max_{k\in {1,2, \dots ,|S|}}  score(i, \theta_k)  \right) -score(i, \theta_j)
    \vspace{-0.2cm}
\end{equation}
where $score(i, \theta_j)$ refers to a ranking metric for the task $\task_i$ using the model obtained training the $j^{th}$ task $\task_j$ in the sequence. 
Looking to $mf(i, \theta_j)$ for all $j$ in the sequence allows observing which tasks have a significant negative transfer impact on $\task_i$ (high value) and which have a low negative impact (low value).

\subsection{Neural ranking models and learning}
\vspace{-0.2cm}
We evaluate two different state-of-the-art neural \texttt{IR} models:\\
$\bullet$ The vanilla Bert\cite{bert2019devlin}(noted \textbf{VBert}) estimating a ranking score based on a linear layer applied on the averaged output of the last layer of the Bert language model.\\
$\bullet$ The \textit{Mono-T5-Ranker}\cite{document2020nogueira} (noted \textbf{MonoT5}) based on  a \textit{T5-base} model fine-tuning and trained to generate a positive/negative token.
%, the ranking score being given by the probability generating the positive token.

\textit{Implementation details:}
All models are trained with \textit{Adam} optimizer \cite{adam2015kingma}, the  optimizer state is not reinitialized for each task of the sequence. Indeed, re-initializing the optimizer will lead to observe a spike in the loss function whether addressing a same or a different task due to the state of \textit{Adam} optimizer parameters.
As previous work in \textit{IR} \cite{studying2021lovon,document2020nogueira,bert2019devlin}, we perform sparse retrieval by re-ranking top-1000 most relevant documents retrieved by the BM25 model \cite{proceedings1992robertson}.

For MonoT5 we start with the \textit{t5-base}\footnote{https://huggingface.co/transformers/model\_doc/t5.html} model with a learning rate of $10{-3}$ and batch size of $16$. For the VBert model\footnote{using bert-base-uncased pretrain}, the batch size is $16$ with a learning rate of $2\times 10^{-5}$ for Bert parameters and $10^{-3}$ for scoring function parameters.

\vspace{-0.2cm}
\section{MSMarco Continual Learning corpus}
\label{section:MS Marco-topics}
\vspace{-0.2cm}
Our continual learning framework is based on learning from a long sequence of tasks. One main difficulty is to create this sequence considering the availability of \textit{IR} datasets. One method would be to build a sequence of datasets of different domains as in \cite{studying2021lovon}, but the number of datasets adapted to neural \textit{IR} (with a sufficiently large number of queries and relevance judgments) is not sufficient for long sequences setting.
We propose to model the task at a lower granularity level, namely topics, instead of the dataset granularity. In what follows, we present our methodology for creating long sequences of topics using the MSMarco dataset.
Once this dataset is validated, it serves as a base for designing controlled settings related to particular \textit{IR} scenarios (all settings and models are open-sourced \footnote{\url{https://github.com/tgeral68/continual_learning_of_long_topic}}).

\subsection{RQ1: Modeling the long topic sequence}
\label{sec:newdatasetContinual}
\vspace{-0.2cm}
To create the long sequence, we consider the MSMarco dataset \cite{NguyenRSGTMD16}. 
Such dataset is based on real users' questions on Bing. Our intuition is that several queries might deal with the same user's interest (e.g., ``what is the largest source of freshwater on earth?'' or ``what is water shortage mitigation''). These groups of queries denote what we call in the remaining paper \textit{topics}. 
To extract topics, we propose a two-step method: extracting clusters from randomly sampled queries and populating those clusters with queries from the whole dataset. We use a similarity  clustering\footnote{https://www.sbert.net/examples/applications/clustering~(fast~clustering)}  based on query representations obtained using the sentence-BERT model \cite{reimers2019sentence}. The clustering is based on a sample of 50,000 randomly picked queries and estimates the similarity cosine distance according to a threshold $t$ to build clusters of a minimum size of $s$. We then populate clusters using other queries from the dataset according to threshold $t$. 
Finally, we produce the sequence of topics by randomly rearranging clusters to avoid bias of cluster size. Another sequencing method might be envisioned for future work, for instance considering a temporal feature by comparing topic trends in real search logs. In practice, the value of the threshold $t$ differs in each step of clustering and populating, leading to the threshold $t_1$ and $t_2$ (with $t_2<t_1$) to obtain clusters of reasonable size to be used for neural models.  Depending on the value of those hyper-parameters ($t_1$, $t_2$, $s$), we obtain three datasets of topic sequences of different sizes (19, 27, and 74), resp. called \textit{MS-TS}, \textit{MS-TM} and \textit{MS-TL} (for small, medium, large).

Statistics of these three topic sequences are described in Table \ref{tab:clustering_dataset}.
To build the train/validation/test sets, we constraint the validation and the test set to be composed of approximately 40 queries by topic. Notice that we do not use the original split as it remains difficult to consider enough testing examples falling into the created topics.

\begin{table}[t]
    \small
    \centering
    \caption{\small Parameters and statistics of the generated dataset and their inter/intra task similarity metric ($c-score$). The intra-score is the mean $c-score$ when comparing a task with itself, and the inter score when comparing different tasks.}
    \begin{tabular}{lccccccc}
        
        \largehline
         Name & $t_1$ & $s$ & $t_2$ & $|\task|$ & \#queries by topics& inter& intra \\
         \mediumhline
         MS-TS & 0.7 & 40 & 0.5 & 19 &  $3,650\pm 1,812$& $ 3.8\%$ & $31.4\%$\\
         MS-TM & 0.75 & 20 & 0.5 & 27 & $3,030\pm 1,723$ & $ 4.1\%$ & $32.1\%$ \\
         MS-TL & 0.75 & 10 & 0.55 & 74 & $1,260\pm 633$& $ 3.3\%$ & $34.6\%$\\
         \mediumhline
         MS-RS & - & - & - & 19 &  $3,650\pm 1,812$&$ 10.3 
         \%$ & $10.2\%$\\
         MS-RM & - & - & - & 27 & $3,030\pm 1,723$&$ 9.9\%$ & $9.8\%$\\
         MS-RL & - & - & - & 74 & $1,260\pm 633$&$ 8.7\%$ &  $8.8\%$ \\
         \mediumhline
    \end{tabular}
    \label{tab:clustering_dataset}
    \vspace{-0.2cm}
\end{table}

\begin{figure}[t]
     \centering
     \begin{subfigure}[b]{0.48\textwidth}
         \centering
         \includegraphics[width=0.8\textwidth]{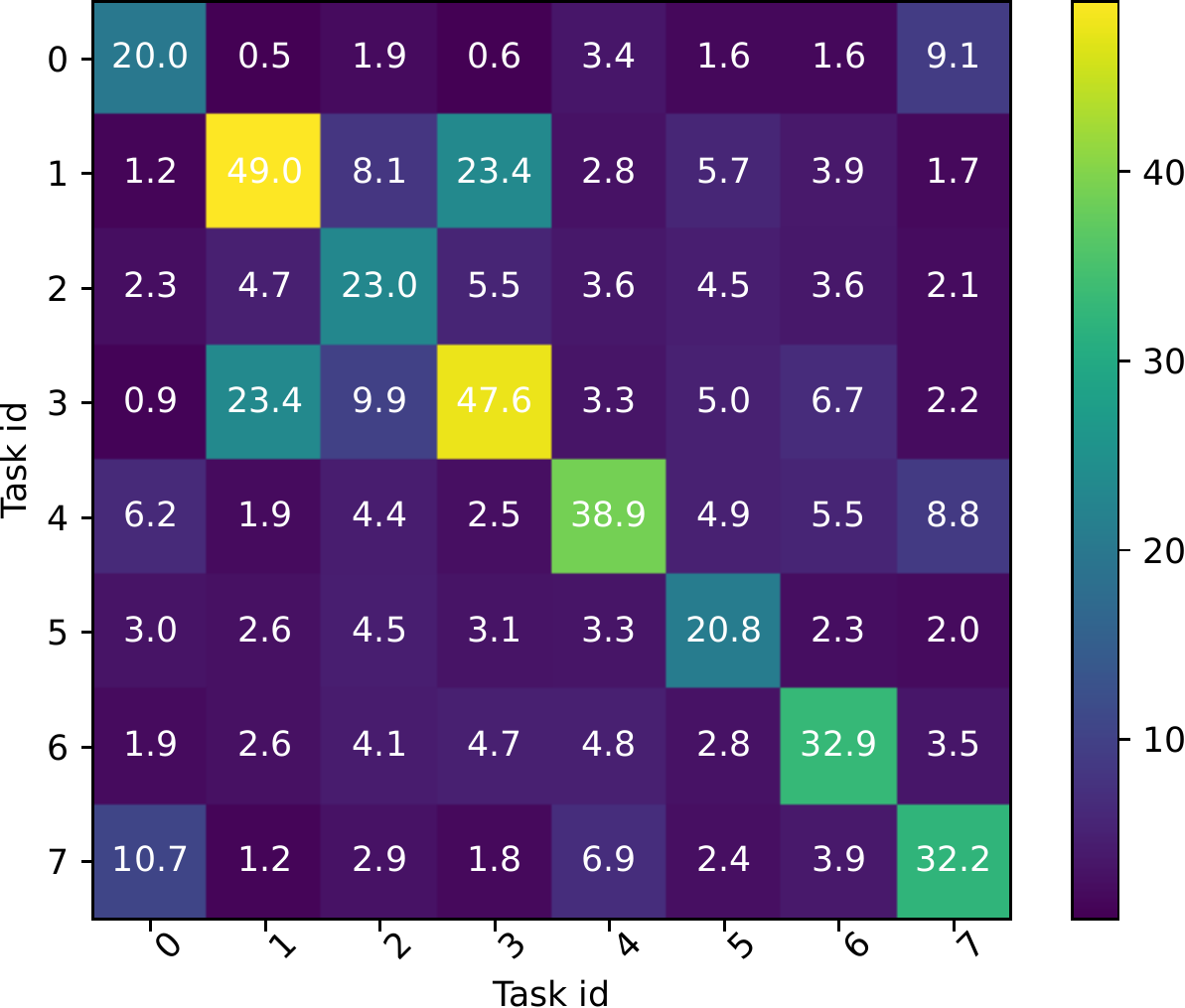}
         \caption{Topics}
         \label{fig:topics}
     \end{subfigure}
     \hfill
     \begin{subfigure}[b]{0.48\textwidth}
         \centering
         \includegraphics[width=0.8\textwidth]{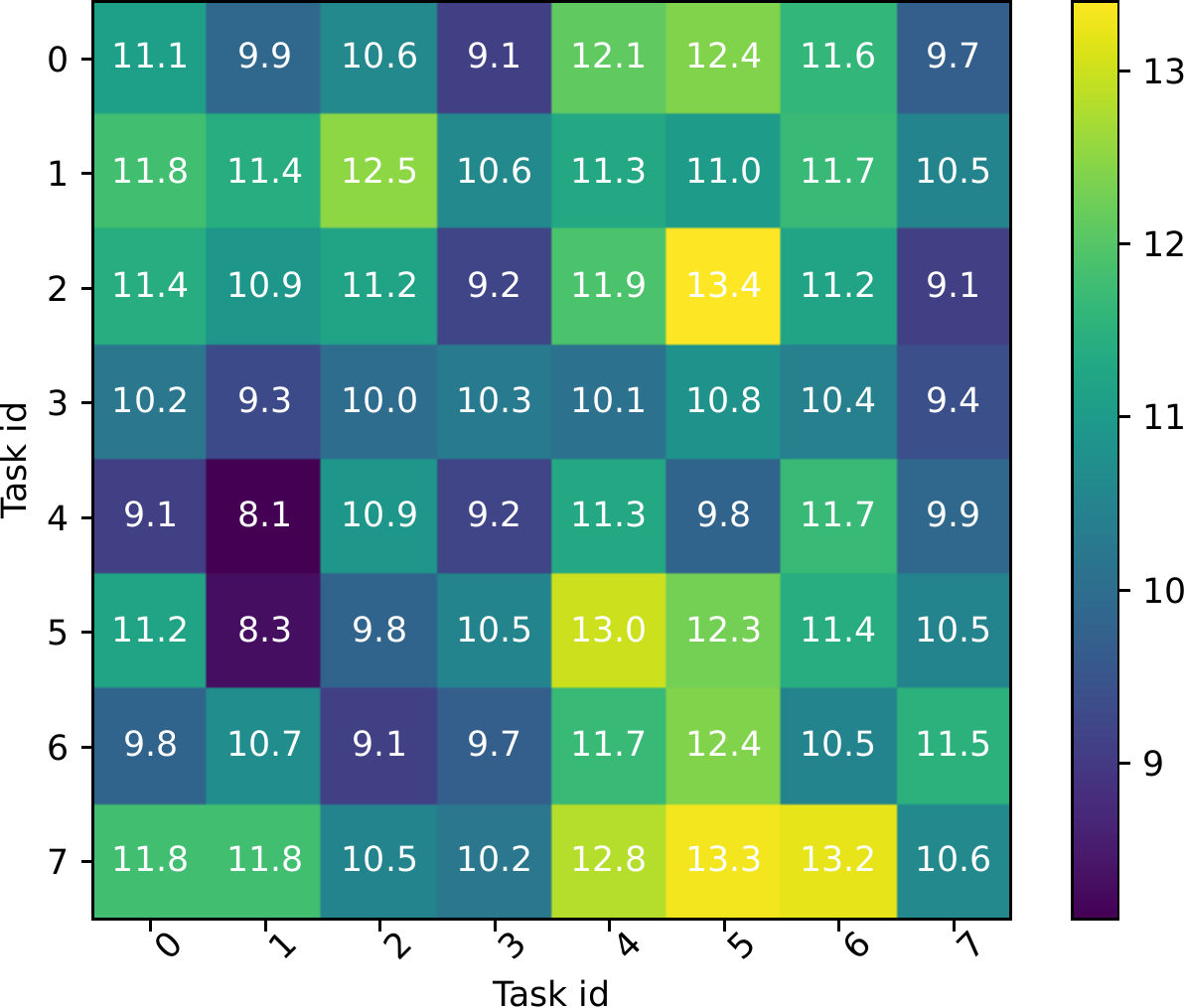}
         \caption{Random}
         \label{fig:random}
     \end{subfigure}
\vspace{-0.3cm}
    \caption{\small Matrix of similarities between topics for 8 tasks of MS-S (\ref{fig:topics}) and MS-RL (\ref{fig:random}) datasets. The c-score ($\times 100$) is processed on all topic pairs, a high value (yellow) denotes the level of retrieved document overlap between queries of topics.}
\vspace{-0.6cm}
    \label{fig:similarity-matrix}
\end{figure}

\vspace{-0.2cm}
\subsection{Evaluating the long topic sequence}
\label{section:topic_evaluation}
\vspace{-0.2cm}

To verify the relevance of the clusters, we aim at measuring retrieval evidence within and between clusters (i.e., queries within clusters might have similar retrieval evidence and queries between clusters might have different ones).  As retrieval evidence, we use the retrieved documents for each query using the  \textit{BM25} model with default parameters \footnote{Implemented in pyserini:  https://github.com/castorini/pyserini}. Our intuition is that similar queries should share retrieved documents (and vice versa). To compare queries within and between clusters, we randomly select two pools (noted $A_{i}$ and $B_{i}$) of 250 queries within each cluster associated to task $\task_i$. Let $D^{A_{i}} = \{D_q| q \in A_{i}\}$ (resp. $D^{B_{i}} = \{D_q| q \in B_{i}\}$) the documents returned by the ranker for the queries in $A_i$ (resp. $B_i$). 

We thus compute the $c-score$ which measures the ratio of common documents between two tasks $\task_i$, $\task_j$ (or same task if $i=j$) as follows:
\vspace{-0.2cm}
\begin{equation}
      c-score(\task_i, \task_j) = \frac{|D^{A_{i}} \cap D^{B_{j}}|}{|D^{A_{i}}|}  
      \vspace{-0.2cm}
\end{equation}
This score is then averaged over pairs of topics within the sequence (intra when comparing topics with their-selves and inter when comparing different topics).

To evaluate our topic sequence methodology, for each of the three datasets we create a long topic sequence baseline in which clusters are extracted randomly from the queries of topics based corpora.
We obtain three randomized datasets denoted \textit{MS-RS}, \textit{MS-RM} and \textit{MS-RL}.

 Table \ref{tab:clustering_dataset}  reports for each of the generated datasets the intra and inter c-scores. By comparing the inter metric between both corpus settings (around 3/4\% for the clustering-based ones and around 9/10\% for randomized ones), one can conclude that our long topic sequence includes clusters that are more different than the ones created in the randomized corpus. The trend is opposite when looking at the intra, meaning that our sequence relies on clusters gathering similar queries but dissimilar from each other. This statement is reinforced in Figure \ref{fig:similarity-matrix} which depicts the $c-score$ matrix for all couples $(i,j) \in \{1,2, \ldots, |S|\}^2$ for a subset of $8$ tasks (for clarity) of the $MS-S$ and $MS-RS$ corpora. % We restricted to 8 tasks for clarity. 
 We observe that for the randomized matrix (Figure 1(b)), the metric value is relatively uniform. In contrast, in the matrix obtained from our long topic sequence based on clustering (Figure 1(a)), the c-score is very small when computed for different topic clusters (low inter similarity) and higher in the diagonal line (high intra similarity).

\vspace{-0.4cm}
\subsection{\textit{IR}-driven controlled stream-based scenario}
\label{sec:stream_def}

In this section, we focus on local peculiarities of the long topic sequence by analyzing \textit{IR}-driven use cases, such as documents or queries distribution shifts. Typically, the available documents may change over time, or even some can be outdated (for instance documents relevant at a certain point in time). Also, it happens that the queries evolve, either by new trends, the emergence of new domains, or shifts in language formulation. To model those scenarios, we propose three different short topic streams to fit the local focus.
Topics are based on our long topic sequence $S=\{\task_1, \ldots, \task_i, \ldots, \task_n\}$ built on MSMarco (Section \ref{sec:newdatasetContinual}). For each scenario, we consider an initial setting $\task_{init}$ modeling the general knowledge before analyzing particular settings. In other words, it constitutes the data used for the pre-training of neural ranking models before fine-tuning on a specific sequence. The proposed controlled settings are:
\vspace{-0.2cm}
\begin{itemize}
    \item \textbf{Direct Transfer \cite{efficient2020veniat}:} The task sequence is   $(\task_{init},\task_i^+, \task_j, \task_i^-)$ where tasks $\task_i^+$ and $\task_i^-$ belong to the topic  task $\task_i$  %$\task_1^+, \task_1^- \subset \task_1$, 
    and have different sizes  ($|\task_i^-| \ll |\task_i^+|$).  This setting  refers to the case when the same topic comes back in the stream with new available data (new queries and new relevant documents).   

    \item \textbf{Information Update:} The task sequence is $(\task_{init},\task_i', \task_i'')$ where $\task_i'$ and $\task_i''$ have dissimilar document distributions and a similar query distribution. Intuitively, it can be interpreted as a shift in the required documents, such as new trends concerning a topic or an update of the document collection. 
    
    \item \textbf{Language Drift:} The task sequence is $(\task_{init},\task_i^*, \task_i^{**})$ where $\task_i^*$ and $\task_i^{**}$ have similar document distributions and a dissimilar query distribution. This can correspond to a change of query formulation or focus in a same topic.
\end{itemize}

\begin{figure}[t]
     \centering
     \begin{subfigure}[b]{0.46\textwidth}
         \centering
         \includegraphics[width=0.75\textwidth]{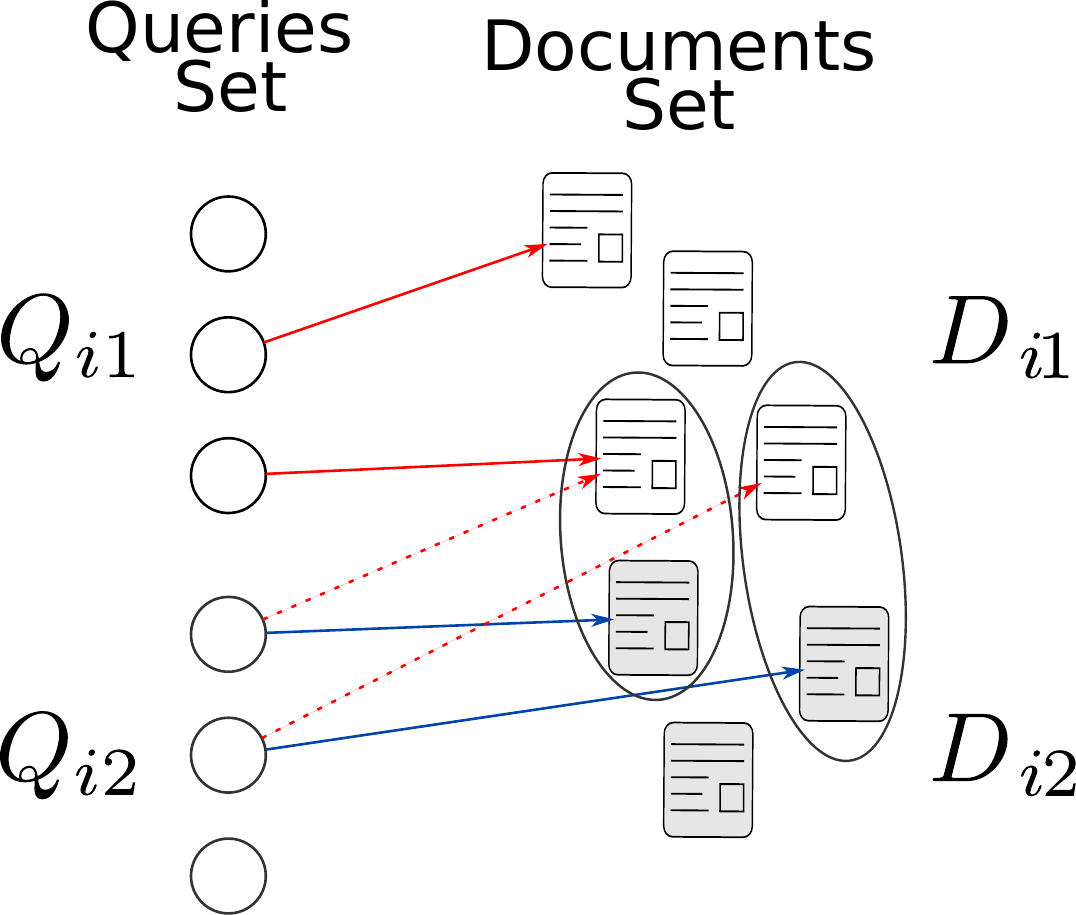}
         \caption{\texttt{IU} scenario}
         \label{fig:IU}
     \end{subfigure}
     \hfill
     \begin{subfigure}[b]{0.45\textwidth}
         \centering
         \includegraphics[width=0.75\textwidth]{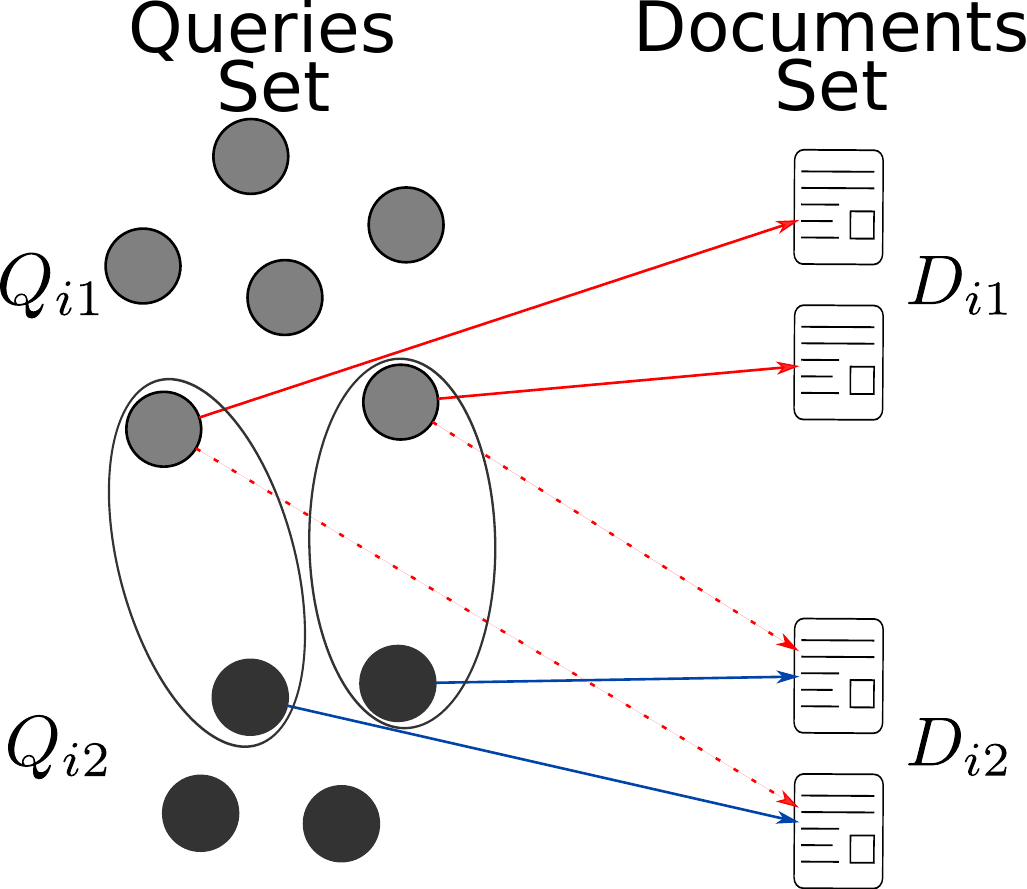}
         \caption{\texttt{LD} scenario}
         \label{fig:LD}
     \end{subfigure}
     \hfill
     \vspace{-0.2cm}
     \caption{\small Both Information Update (IU) and Language Drift (LD) scenarios. The circle of documents or queries represent the pair of documents or queries of different clusters, mapped using the  closest neighborhood algorithm. This mapping is used to infer query-relevant documents of different clusters (dotted lines). Solid lines correspond to original query-relevant documents pairs.  The red arrows build the training sets of tasks  $\task'$ and $\task^*$  while blue arrows compose the one of tasks $\task''$ and $\task^{**}$. }
     \vspace{-0.4cm}
\end{figure}

To build those sequences, the initial task $\task_{init}$ aggregates $k$ different tasks available in the original sequence topics $S$. We set $k=5$ which is a good balance between considering enough tasks for the pre-training and considering not too many tasks to allow an impact of model fine-tuning on our controlled settings. 

For the \textbf{Direct Transfer}, we randomly select a set of three topics (metrics are then averaged),  $75\%$ of the queries are used for $\task_i^+$ and $25\%$ for $\task_i^-$. $\task_j$ is a topic selected randomly.

For \textbf{Information Update}, we consider that, for persistent queries, relevant documents might evolve. To do so, we randomly select  three topics $\task_i$.
For each topic $\task_i$, we cluster the associated relevant documents using a constrained 2-means algorithm\footnote{https://pypi.org/project/k-means-constrained/} based on the cosine similarity metric of Sentence Bert embeddings (used in the section \ref{sec:newdatasetContinual}). We obtain two document sets $D_{i1}$ and $D_{i2}$: the initial and final information distribution. Since queries in MSMarco passages have in a vast majority one relevant document\footnote{if not the case, we sample one document to build the query-relevant document pairs}, we can easily obtain the set of queries $Q_{i1}$ and $Q_{i2}$ associated to document sets $D_{i1}$ and $D_{i2}$ (see Figure 2(a) - solid lines being the query-document relevance pairs).  
To model the information update, we map  documents $D_{i2}$ relevant for queries in $Q_{i2}$ (final distribution) to most similar documents in $D_{i1}$ (initial distribution) in the embedding space (circles in Figure 2(a)).
The task $\task_i'$ considers the whole set of queries $Q_{i1}$ and $Q_{i2}$ but only the document set $D_{i1}$ as initial information (red arrows in Figure 2(a)). The task $\task_i''$ corresponds to the update of the information (namely, documents). We thus only consider the query set $Q_{i2}$ for persistent queries with the document set $D_{i2}$ as information update (blue arrows in Figure 2(a)). 
We expect that $Q_{i1}$ performs similarly after information update if models do not suffer from catastrophic forgetting and that $Q_{i2}$ improves its performance with the information update.  We also consider the reversed setting in which we first consider  $D_{i2}$ as the initial information and then update the information with $D_{i1}$, $Q_{i1}$ (persistent queries). 

For the \textbf{Language Drift} scenario, we use a similar protocol by clustering queries instead of documents to obtain the sets of queries $Q_{i1}$ and $Q_{i2}$, and then the associated relevant document sets $D_{i1}$ and $D_{i2}$. To model the language drift in queries, we consider that one query set will change its query formulation. To do so, let consider that sets $Q_{i1}$ and $Q_{i2}$ reflect resp. the initial and final language distribution of same information needs, and thus, requiring same/similar relevant documents.
To observe the language drift, we  map pairs of queries $(q_{i1},q_{i2}) \in Q_{i1} \times Q_{i2}$  according to their similarity in the embedding space (circles in Figure 2(b)). 
Thus, we can associate documents of $D_{i2}$ (document relevant for queries of $Q_{i2}$) to the query set $Q_{i1}$: $q_{i1}$ has two relevant documents ($d_{i1}$ and $d_{i2}$) (red arrows in Figure 2(b)). 
The $\task_i^{*}$ is composed of the query set $Q_{i1}$ and the associated relevant documents belong to both $D_{i1}$ and $D_{i2}$ (red arrows).
The $\task_i^{**}$ is based on the query set $Q_{i2}$ (new language for similar information needs) associated to the relevant documents $D_{i2}$ (blue arrows). We also consider the reversed setting in which query sets $Q_{i2}$ and $Q_{i1}$ are resp. used for the initial and final language. \\
For those two last scenarios (information update and language drift), metrics are respectively averaged over initial and reversed settings.

\vspace{-0.2cm}
\section{Model performance and learning behavior on long topic sequences}
\label{section:experiments}
\vspace{-0.2cm}
In this section, we report the experiments on the continual settings proposed in Section \ref{section:MS Marco-topics}. We first analyze the overall retrieval performance of the different models applied on long topic sequences.  We then present a fine-grained analysis of the different models with a particular focus on catastrophic forgetting regarding the similarity of topics in the sequence. Finally, we analyze specific \textit{IR} use cases through our controlled settings. 

\begin{figure}[t]
\begin{subfigure}{.58\textwidth}
\begin{scriptsize}
    
    \begin{tabular}{ccccc}
    \largehline
         Model &  Dataset  & \multicolumn{3}{c}{Learning protocol} \\
         & & Random &  clustering & Multi-task\\
         \mediumhline
         \multirow{3}{*}{VBert}    & SMALL & $18.4/19.6$& $16.3/17.5$ & $\mathbf{18.5}/\mathbf{19.7}$\\
                                            & MEDIUM & $\mathbf{17.9/19.0}$  & $17.8/18.9$ & $17.5/18.7$ \\
                                            & LARGE &  $\mathbf{18.8/19.9}$ & $17.3/18.5$ & $18.5/19.7$\\
                                        \hline
         \multirow{3}{*}{MonoT5}     & SMALL & $\mathbf{16.1}/\mathbf{17.3}$ & $13.1/14.4$& $15.5/16.8$ \\
                                        & MEDIUM   & $15.4/16.7$  & $13.4/14.7$ & $\mathbf{15.7}/\mathbf{17.1}$\\
                                        & LARGE  & $13.9/15.1$ & $13.8/15.1$ & $\mathbf{15.7}/\mathbf{17.0}$\\
        \mediumhline
         \multirow{3}{*}{BM25}            & SMALL&  \multicolumn{3}{c}{$10.8/11.7$}\\
                                            & MEDIUM & \multicolumn{3}{c}{$10.5/11.4$} \\
                                            &LARGE & \multicolumn{3}{c}{$11.7/12.7$}\\
        \mediumhline
         
    \end{tabular}
    \end{scriptsize}
    \caption{Mean performances on all the tasks reporting $mrr@10/mrr@100$ for the different models.}
         \label{tab:perfomance_by_task}
    \end{subfigure}
    \begin{subfigure}{.40\textwidth}
        \includegraphics[width=1.1\linewidth
    ]{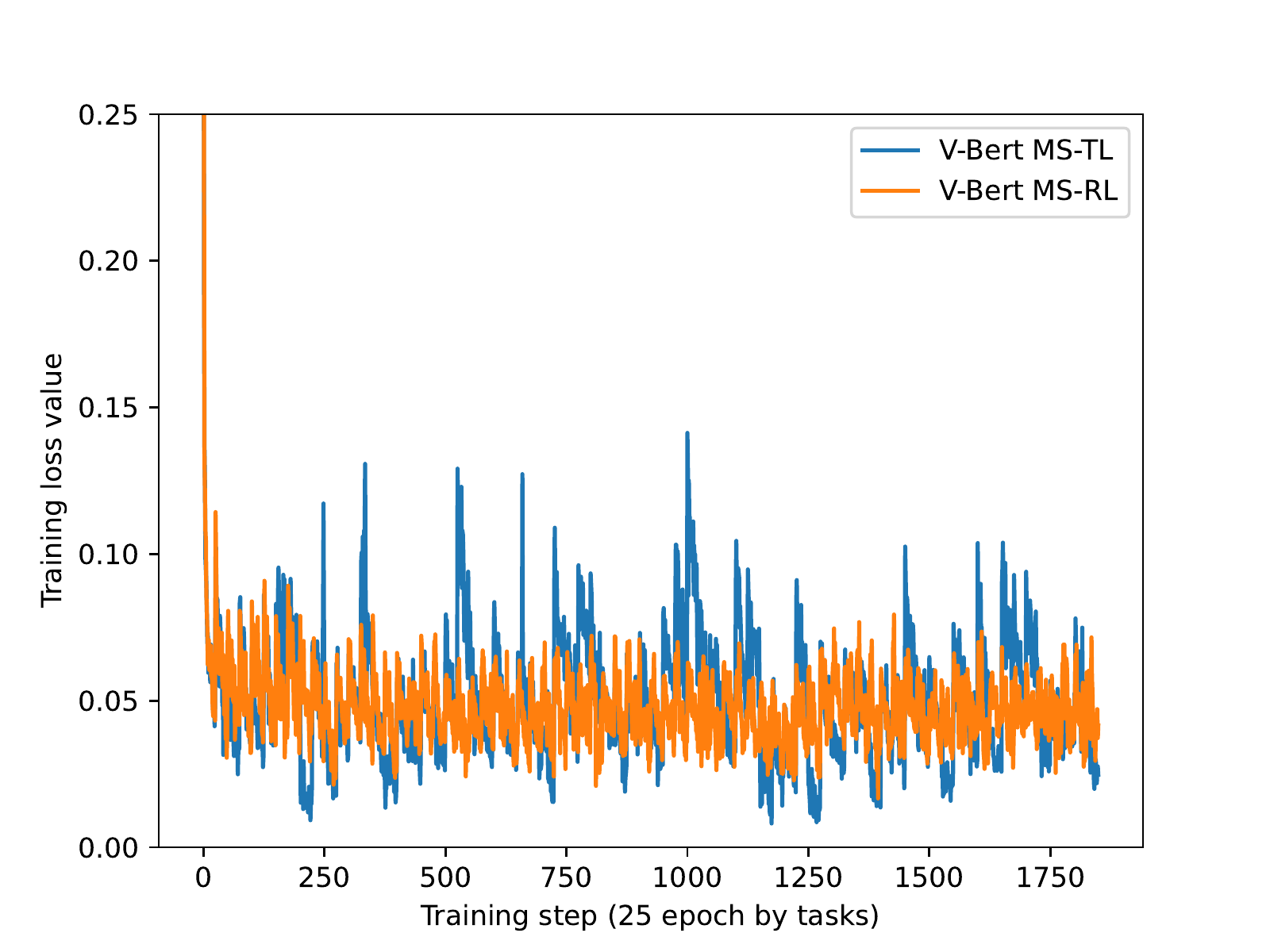}
        \caption{VBert loss values for both random and clustering-based large corpus.}
        \label{fig:loss_bert}
    \end{subfigure}
    \vspace{-0.2cm}
    \caption{General performance of neural ranking models on the long topic sequence. }
    \vspace{-0.4cm}
\end{figure}

\vspace{-0.2cm}
\subsection{RQ2: Performances on the MSMarco long topic sequence}
\vspace{-0.2cm}
We focus here on the global performance of neural ranking models after having successively been fine-tuned on topics in our MSMarco-based long sequence setting (Table \ref{tab:perfomance_by_task}). For comparison, we use different sequence settings (i.e., the randomized and the topic clustering ones) of different sizes (i.e., small, medium, and large). We also run the multi-task baseline in which models are trained on all the tasks of the sequence jointly (without sequence consideration). 
At a first glance, we can remark that, in a large majority, neural models after fine-tuning on random sequences or multi-task learning obtain better results than after the fine-tuning on our long topic sequences. This can be explained by the fact that, within our setting, the topic-driven sequence impacts the learning performance: a supplementary effort is needed by the model to adapt to new domains, which is not the case in the random setting. In this latter, the diversity is at the instance level. This trend is depicted in Figure \ref{fig:loss_bert}, highlighting peaks in the clustering-based setting (blue line) referring to topic/cluster changes. This result confirms that catastrophic forgetting might occur with neural ranking models. 
\vspace{-0.3cm}
\subsection{Fine-grained analysis}
\label{section:label_transfer}

\begin{figure}[t]
\begin{subfigure}{.58\textwidth}
    \centering

\scriptsize
    \begin{tabular}{cccccc}
        \largehline
         Model & Dataset &  $1^{st}$& $2^{nd}$ & $3^{rd}$ & $4^{th}$ \\
         \mediumhline
         \multirow{2}{*}{Mean Similarity by quartile}    & MS-S & 1.4 & 2.6& 3.9 & 13.8 \\
             & MS-M & 1.5 & 2.8& 4.7 & 15.3 \\
         
         \mediumhline
         \mediumhline
         \multirow{2}{*}{VBert} & MS-S & 6.3& 6.4& 5.4& \textbf{4.6}\\
         & MS-M  & 4.2& 4.4& 5.1& \textbf{3.8}\\
         \hline

         \multirow{2}{*}{MonoT5} & MS-S& 9.2& 7.0& 6.5& \textbf{6.3}\\
         & MS-M & 6.5& 5.3& 6.0& \textbf{4.5}\\

         \largehline
    \end{tabular}
        \caption{Mean $mf$ score grouped by similarities between tasks (mean of 5 selected topics). The results are averaged according to quartile based on the task similarity metric. The mean value of grouped similarity are reported in the head of the table.}
\vspace{-0.5cm}
    \label{tab:transfer}
\end{subfigure}
\hspace{0.1cm}
\begin{subfigure}{.4\textwidth}
    \centering
    \includegraphics[width=1\linewidth]{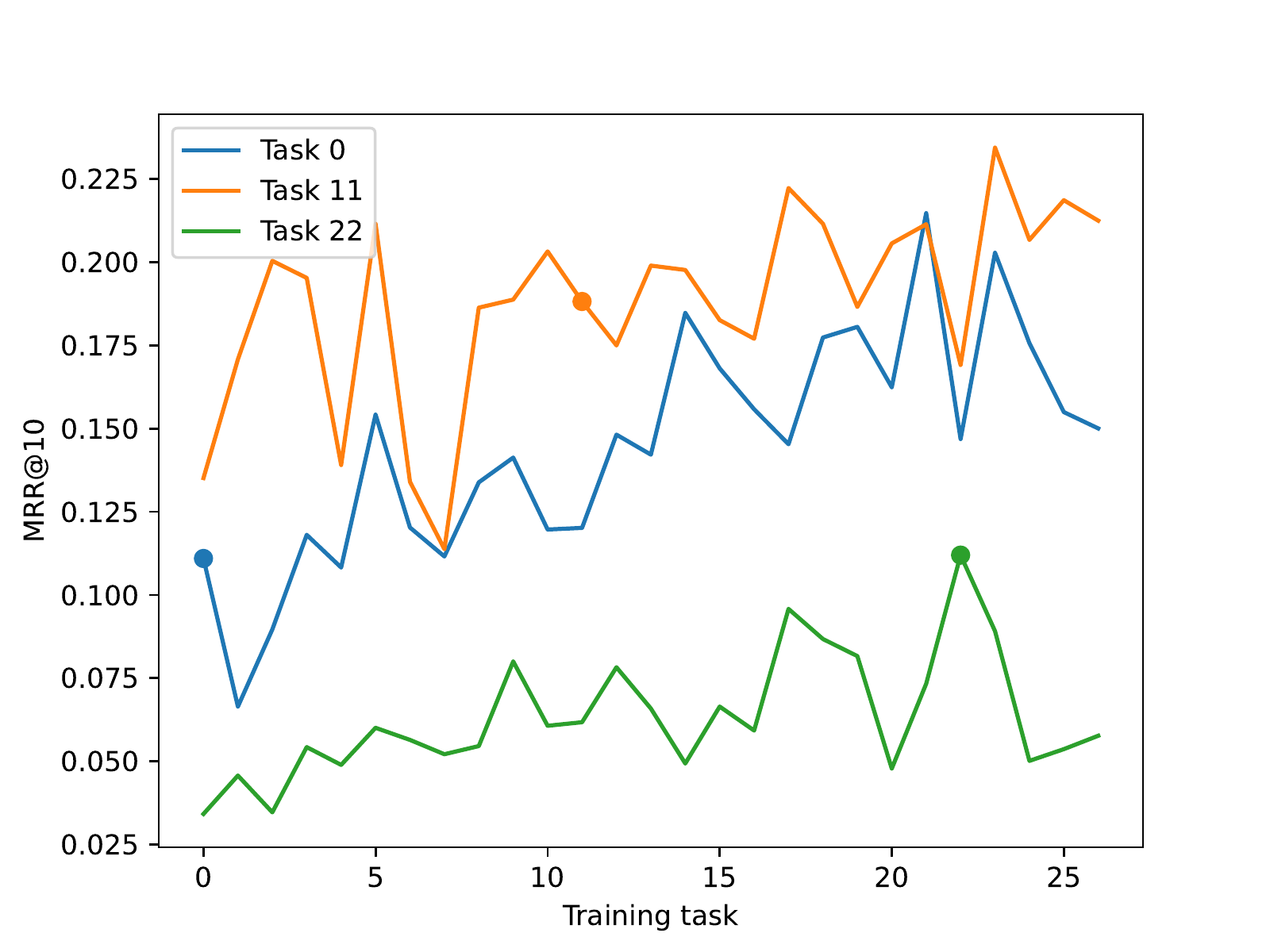}

    \caption{MRR@10 results for three tasks tracked along the training sequence}

    \label{fig:mrr10_long_seq}

\end{subfigure}
\caption{\small Fine-grained analysis of neural ranking model in the long topic sequence.}
\vspace{-0.4cm}
\end{figure}

To get a deeper understanding of model behavior, we aim here to analyze the model performance throughout the learning of the sequence. We are particularly interested in explaining the possible behavior of catastrophic forgetting according to the similarity level between tasks in the sequence. For computational reasons, we were not able to track all tasks throughout the whole sequence, we thus considered 5 randomly selected tasks (as described  in Section 3.1). For each of these 5 tasks $\task_i$, we estimate the catastrophic forgetting using the $mf$ score (Equation 1) regarding each task $\task_j$ of the sequence (with $i \neq j$). 
For the similarity metric, we use the $c-score$ (Equation 2) computed between both tasks $\task_i$ and $\task_j$. 
In Table \ref{tab:transfer}, we group together similarity by quartiles and estimate the average of the $mf$ score for tracked tasks in each similarity quartile. We first remark that the mean similarity values of quartiles are relatively small (except the $4^{th}$ quartile), reinforcing the validation of our dataset building methodology. Also, we observe the following general trends. First, neural ranking models suffer from catastrophic forgetting (positive $mf$ score), particularly the MonoT5 model. The difference in terms of model on both the global effectiveness (Figure 3(a)) and the similarity analysis suggests that MonoT5 is more sensible to new domains than the VBert model. This can also explain by the difference in the way of updating weights (suggested in the original papers \cite{bert2019devlin,document2020nogueira}). In VBert, two learning rates are used: a small one for the Bert model and a larger for the scorer layer; implying that the gradient descent mainly impacts the scorer. In contrast, the MonoT5 is learnt using a single learning rate leading to modify the whole model. 
 Second,  more tasks are similar (high $c-score$), less neural ranking models forget (low $mf$). In contrast to continual learning in other application domains \cite{overcoming2017kirpatrick,icarl2017rebuffi} in which fine-tuning models on other tasks always deteriorates task performance, our analysis suggests that tasks might help each other (particularly when they are relatively similar), at least in lowering the catastrophic forgetting. 
Moreover, as discussed in  \cite{GuoFAC16}, relevance matching signals play an important role in the model performance, often more than semantic signals. The task sequence may lead to a synergic effect to perceive these relevance signals.
Figure \ref{fig:mrr10_long_seq} shows the VBert performance for three tasks located at different places in the sequence (circle point). To perceive catastrophic forgetting, we look at one part of the curve after the point. One can see that task performances increase after their fine-tuning (higher increase when the task is at the beginning of the sequence), highlighting this synergic effect. 
In brief, continual learning in \textit{IR} differs from usual classification/generation lifelong learning setting. It is more likely to have different tasks allowing to ``help'' each other, either by having closely related topics or by learning a similar structure in the query-document matching. 

\vspace{-0.2cm}
\subsection{RQ3: Behavior on \textit{IR}-driven controlled settings}

In this section we review the different scenario described in the section \ref{sec:stream_def}: \textbf{Direct Transfer} (\texttt{DT}), \textbf{Information Update} (\texttt{IU}) and \textbf{Language Drift} (\texttt{LD}).
For all the different settings, we estimate the average metric of the different tracked tasks after each sequence step.

Table \ref{tab:scenarios:dt} reports the effectiveness of neural models on  task $\task_i$ ($\task_i^+$ and $\task_i^-$ being subsets of $\task_i)$ after each fine-tuning step in the \textbf{Direct Transfer scenario}. One can see that fine-tuning on a foreign domain ($\task_2$), the performance of both models on task $\task_i$ drop, highlighting a behavior towards catastrophic forgetting. However, both models are able to slightly adapt their retrieval performance after the fine-tuning of task $\task_i^-$. This final performance is however lower than the baseline model (training on both $\task_{init}$ and $\task_i$) and for the VBert model lower than its initial performance in the beginning of the learning sequence. These two last statements suggest the ability of neural models to quickly reinject a part of the retrained knowledge learnt in the early sequence to adapt to new query/document distributions in the same topic.

\begin{figure}[t]
    \begin{subfigure}{0.35\linewidth}
        
    \centering
    \footnotesize
    \begin{tabular}{c|c|c|c|c}
 
         & \multicolumn{3}{c|}{DT scenario}& B \\
        \hline
         & $\task_i^+$ & $\task_j$ & $\task_i^-$  \\

        \hline
        MonoT5&$26.6$&$24.9$&$26.6$& $27.2$  \\

        VBert& $28.5$ & $26.7$ & $27.3$ & $28.9$\\
        \hline
    \end{tabular}
     \caption{MRR@10 for task $\task_i$ in the Direct Transfer (DT) scenario. See Section 4.3 for building $\task_i$ and $\task_j$.}
     \label{tab:scenarios:dt}
    \end{subfigure}
    \hspace{0.2cm}
    \begin{subfigure}{0.6\linewidth}
        
    \centering

\footnotesize
    \begin{tabular}{cc|c|c|c||c|c|c}

    && \multicolumn{3}{c||}{IU} &  \multicolumn{3}{c}{LD}\\
    &  & $\task'_i$ & $\task''_i$  & B &$\task^{*}_i$ & $\task^{**}_i$ & B\\
        \largehline

        \parbox[t]{1.6mm}{\multirow{3}{*}{\rotatebox[origin=c]{90}{\scriptsize{MonotT5}}}}&$Q_{i1}D_{i1}$ & $28.15$ & $29.6$ &-& $15.6$ & $23.0$ & - \\
        &$Q_{i2} D_{i2}$ & $7.75$ & $26.0$  &-& $16.8$ & $26.5$& -\\

        &$Q_{i1} D_{i1} \cup Q_{i2} D_{i2} $& $18.2$ & $27.8$ & 27.2 & $15.6$ & $23.8$ & 27.2\\
        \mediumhline

        \parbox[t]{1.6mm}{\multirow{3}{*}{\rotatebox[origin=c]{90}{\scriptsize{VBert}}}}&$Q_{i1} D_{i1}$ & $23.7$ & $30.2$ & - & $28.2$ & $30.1$ & -\\
        &$Q_{i2} D_{i2}$ & $14.5$ & $31.4$ & - & $25.5$ & $25.5$ & -\\

        &$Q_{i1}D_{i1} \cup Q_{i2} D_{i2}$ & $19.1$ & $30.9$ & $28.9$  & $26.6$ & $27.0$ & $28.9$\\
        
        \largehline

    \end{tabular}

 \caption{MRR@10 for the Information Update (IU) and Language Drift (LD)  scenarios. See Section 4.3 for the explanation of sets. }
    \label{tab:scenarios:results}
    \end{subfigure}
    \vspace{-0.2cm}
    \caption{\small Model performances on \textit{IR}-driven controlled settings. B stands for the baseline.}
    \vspace{-0.4cm}
\end{figure}

Table \ref{tab:scenarios:results} reports the average effectiveness metrics for both \textbf{Information Update} (\texttt{IU}) and \textbf{Language Drift} (\texttt{LD}) scenarios on different sets, $Q_{ik}D_{i}$ (k=1,2) denoting the sets used to build relevant pairs of query-document (see Section \ref{sec:stream_def}). 
In \texttt{IU} scenario, relevant documents of certain queries ($Q_{i2}$) evolve over time ($D_{i1} \rightarrow D_{i2}$). For both $Q_{i1}D_{i1}$ and particularly $Q_{i2}D_{i2}$ whose queries have encountered the information update, evaluation performances increase throughout the fine-tuning process over the sequence. This denotes the ability of models to adapt to new document distributions (i.e., new information in documents). The adaptation is more important for the MonoT5 model ($7.75$ vs. $26.0$ for the $Q_{i2}D_{i2}$ set), probably explained by its better adaptability to new tasks (as discussed in section \ref{section:label_transfer}). Interestingly, the performance at the end of the learning sequence overpasses the result of the baseline (fine-tuning on $\task_i$): contrary to the direct transfer scenario, this setting has introduced pseudo-relevant documents in task $\task'_i$ which might help in perceiving relevance signals.

For the \textbf{Language Drift }\texttt{LD} scenario, the behavior is relatively similar in terms of adaptation: performances increase throughout the sequence and MonoT5 seems more flexible in terms of adaptation. However, it seems more difficult to sufficiently acquire knowledge to reach the baseline performance (although pseudo-relevant documents have also been introduced). This might be due to the length of queries, concerned by the distribution drift: when the vocabulary changes in a short text (i.e., queries), it is more difficult to capture the semantics for the model and to adapt itself in terms of knowledge retention than when the change is carried out on long texts (i.e., documents as in the information update).

\vspace{-0.4cm}
\section{Conclusion and future work}
\label{section:conclusion}
\vspace{-0.3cm}
In this paper, we proposed a framework for continual learning based on long topic sequences and carried out a fined-grained evaluation, observing a catastrophic forgetting metric in regards to topic similarity.
We also provided specific stream of tasks, each of them addressing a likely scenario in case of \textit{IR} continual learning. Our analysis suggests different design implications for future work: 1) catastrophic forgetting in \textit{IR} exists but is low compared to other domains \cite{overcoming2017kirpatrick,efficient2020veniat}, 2)  when designing lifelong learning strategy, it is important to care of task similarity, the place of the task in the learning process and of the type of the distribution that needs to be transfered (short vs. long texts).
We are aware that results are limited to the experimented models and settings and that much remains to be accomplish for more generalizable results. But, we believe that our in-depth analysis of topic similarity and the controlled settings is a step forward into the understanding of continual \textit{IR} model learning.

\paragraph{\textbf{Acknowledgements.}} We thank the ANR JCJC SESAMS project (ANR-18-CE23-0001) for supporting this work. This work was performed using HPC resources from GENCI-IDRIS (Grant 2021-101681).
%This work was performed using HPC resources from GENCI-IDRIS (Grant 2021-101681) 
\bibliographystyle{spmpsci}
\bibliography{continual-ranking}

\end{sloppypar}

\end{document}